# Enhanced spin lifetimes in a two dimensional electron gas in a gate-controlled GaAs quantum well

Sergiu Anghel[1], Akshay Singh[2], Felix Passmann[1], Hikaru Iwata[3], Nick Moore[3], Go Yusa[3], Xiaoqin Li[2,4], Markus Betz[1]

[1] Experimentelle Physik 2, Technische Universität Dortmund, Otto-Hahn-Straße 4a, D-44227 Dortmund, Germany
[2] Department of Physics and CQS, University of Texas at Austin, Austin, Texas 78712, USA
[3] Department of Physics, Tohoku University, Sendai 980-8578, Japan
[4] Texas Materials Institute, University of Texas at Austin, Austin, Texas 78712, USA
E-mail address: markus.betz@tu-dortmund.de

Exciton, trion and electron spin dynamics in a 20 nm wide modulation-doped GaAs single quantum well are investigated using resonant ultrafast two-color Kerr rotation spectroscopy. Excitons and trions are selectively detected by resonant probe pulses while their relative spectral weight is controlled by adjusting the gate voltage which tunes the carrier density. Tuning the carrier density markedly influences the spin decay time of the two dimensional electron gas. The spin decay time can be enhanced by a factor of 3 at an intermediate carrier concentration in the quantum well, where excitons and trions coexist in the system. In addition, we explore the capability to tune the g-factor of the electron gas via the carrier density.

## I. INTRODUCTION

A two-dimensional electron gas (2DEG) is a semi-metallic sea of electrons in a quantum well (QW). It is typically created by doping and controlled via a back-gate voltage. The investigation of spin dynamics in a 2DEG is technologically important as the electrons often have relatively long spin lifetimes in the nanosecond range, useful for spintronic applications [1-3]. Upon optical excitation, excitons (or electron-hole pairs) are created, which can capture an extra charge to form trions [4-6]. In negatively charged trions, two electrons form a spin singlet state. After the trion recombines, the electron left behind maintain its spin orientation or coherence [7-9]. Thus, trions offer an optical means to initialize and control the coherent spin polarization of the 2DEG. A careful study of the 2DEG spin dynamics in the presence of optically excited quasiparticles (excitons and trions) is of great interest for optically controlled spin devices [3,10].

Many previous experiments exploring this interplay of 2DEG and optical quasiparticles, were performed in II-VI QWs (e.g. CdTe) where trion binding energy is relatively large [8,11,12]. However, the large spin-orbit coupling in II-VI QWs also leads to a faster spin dephasing in comparison to the III-V QWs (e.g. GaAs) [3,13]. In GaAs QWs, trions have a small binding energy of a few meV. Thus, well-resolved exciton and trion resonances can only be observable in the highest quality QWs [14]. Here, we focus on modulation doped wide QWs with thickness comparable to the excitonic Bohr radius. These wide QWs' are less susceptible to monolayer fluctuations which are commonly seen in narrower QWs', and which limit carrier mobility due to disorder induced localization. In addition, modulation doped QWs' provide gate dependent tuning of the electron (or hole) density in the QW, giving insight into phenomena dependent on the relative exciton-trion density [10,14-16]. Remarkably, wide QWs' are characterized by longer spin lifetimes for excitons and 2DEG when compared to narrow QWs. This finding is related to reduced spin scattering [17]. Compared to bulk materials, spins in wide QWs have reduced lifetimes, but can be effectively initialized through trions [18]. A combination of fast modulation through excitonic quasiparticles, and storage of coherence through the 2DEG makes wide QWs promising for logic and memory applications.

In this Article, we investigate spin dynamics using Kerr rotation (KR) spectroscopy in a high quality, 20 nm wide, modulation doped GaAs QW. The electron mobility exceeds $10^6$ cm$^2$/Vs at 4 K. We investigate spin lifetime of excitons, trions and the 2DEG and their dependence on the electron density in the QW tunable via the back-gate voltage. We observed a markedly enhanced 2DEG spin lifetime in the voltage range where the QW features a transition from an insulating to a conducting state. In the optical response, this regime is characterized by a balanced coexistence of excitons and trions. In addition, an analysis of the magneto-optic Kerr effect (MOKE) in Voigt geometry provides the transverse electronic Landé g-factors of the 2DEG. The variation of g-factor with the back-gate voltage reveals a modification in carrier localization [12].

## II. EXPERIMENTAL DETAILS

The QW structure (Fig. 1a) is grown by molecular beam epitaxy on a (100) oriented, highly n-doped GaAs substrate. The growth starts with 50 periods of AlGaAs/GaAs (10nm/3nm) to reduce charge leakage from the substrate. The 20 nm wide GaAs QW of interest, is sandwiched between two $Al_{0.33}Ga_{0.67}As$ layers. A Si-delta doping growth interruption with a nominal concentration of n ~ $4 \times 10^{12}$ cm$^{-2}$, located 96 nm away from the QW, provides n-doping and allows for electrical tuning of the carrier density in the QW. The sample is processed into a Hall-bar with Ohmic AuGeNi contacts, and features a back-gate used for carrier density tuning.

The sample is mounted in a very stable helium flow cryostat (Oxford Microstat HiRes2). It is cooled down to a temperature of ~ 3.5 K in order to ensure narrow exciton and trion resonances and, potentially, long spin lifetimes. We use non-degenerate KR in confocal reflection geometry to study the spin dynamics in the QW [19]. To this end, the output of a mode-locked femtosecond Ti-sapphire laser of 80 MHz repetition rate is split into two paths. Both pulse trains pass independent grating based 4f-pulse-shapers for wavelength tuning [20]. The temporal (spectral) resolution of the system is 1 ps (0.7 nm). For all results presented below, the pump beam is tuned to ~1.55 eV (800 nm) to effectively initialize spin. The polarization state of the pump pulse is modulated between $\sigma^+$ and $\sigma^-$ helicities. The linearly polarized probe beam is tuned into resonance with either the exciton or the trion transition. Both beams are collinearly focused onto the sample using a Mitutoyo M-Plan APO NIR 50X objective. We operate with a slightly defocussed spot of ~8 $\mu$m, for both pump and probe, to limit the density of photo-excited carriers and to increase the signal-to-noise ratio. Residual pump light is filtered out by a spectrometer. The KR signal is detected in a standard balanced detection scheme. The amplitude of the signal is proportional to the rotation of the probe polarization associated with the pump induced initialization of spins. KR is a particularly sensitive technique to probe long lived coherences in QWs', and is, hence, an ideal tool to study spin phenomena in 2DEGs [19]. Spin dynamics are measured as a function of delay time between pump and probe pulses. To investigate the MOKE response, a magnetic field generated by an electromagnet is applied in the Voigt geometry.

## III. RESULTS AND DISCUSSION

We start the sample characterization by analysing the photoluminescence (PL) of the QW as a function of the back-gate voltage. Specifically, the sample is excited by ~1.55 eV (800 nm) light from the Ti:sapphire laser, with intensities similar to the one used as a pump for the time-resolved KR measurements below (40 $\mu$W). The PL spectra in Fig. 1(b) map the exciton (X) and trion (T) resonances, as well as their relative spectral weights with changes of the electron density in the QW. The electron density in the QW increases as the back-gate voltage is tuned towards more positive values. Around $V_B$ ~ -0.5 V, the exciton and trion features are spectrally clearly separated. In contrast, below a back-gate voltage of $V_B$ ~ -1 V, the exciton and trion resonances can no longer be well resolved and start to merge. The energy separation between the trion and exciton resonances (i.e. the trion binding energy) changes systematically with the carrier density [14,21,22]. Beyond $V_B$ ~ +0.5 V, the exciton and trion features start to vanish. The spectral features evolve into a broader background signal reflecting radiative transitions in denser 2DEG. As detailed below, the voltage regime of – 0.5 V < $V_B$ < 0.5 V with its simultaneous presence of excitons and trions is most interesting in terms of enhanced spin lifetimes. To selectively address excitons and trions in the KR measurements, the probe photon energies are tuned to 1523 meV (X) and 1517.5 meV (T) for excitons and trions, respectively.

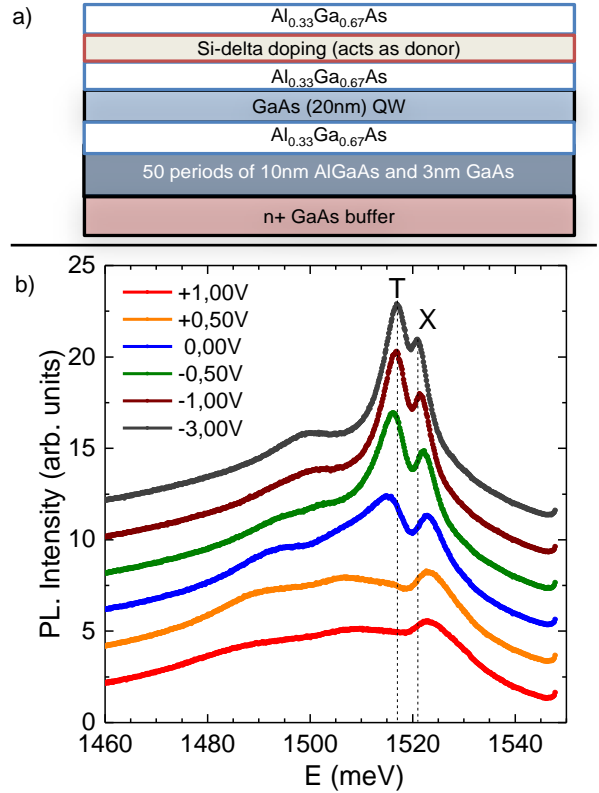

FIG. 1: a) Schematic of the heterostructure. b) Photoluminescence (PL) for various back-gate voltages. The curves are vertically shifted for clarity. The trion (exciton) resonance is labeled with T(X).

To measure the relevant spin relaxation timescales of excitons, trions and the 2DEG, we now carry out time-resolved KR. The pump is again fixed at 1550 meV, and the probe is selectively tuned to exciton (X, 1523 meV) and trion (T, 1517.5 meV) transitions. We first consider a gate voltage of $V_B$ = 0 V. To enhance signal-to-noise ratio and to eliminate background signals, we extract the difference between signals for $\sigma-$ (left circularly polarized) and $\sigma+$ (right circularly polarized) pump helicities. The resulting KR transients for exciton and trion are shown in Fig. 2. The dynamics can be described as a bi-exponential decay. While we attribute the fast component to the quasiparticle (exciton, trion) spin relaxation, the long-lived dynamics is related to spin polarization of the 2DEG. For a more quantitative evaluation, the differential pump-probe signals ($dR/R$) are fitted to a bi-exponential decay

$$\frac{dR}{R} = \left( A_1 e^{-\frac{t-t_0}{\tau_1}} + A_2 e^{-\frac{t-t_0}{T_1^{2DEG}}} \right) \quad (1),$$

where $A_1(A_2)$ and $\tau_1(T_1^{2DEG})$ are the amplitudes and spin lifetimes for the short-lived (long-lived) component, respectively. The fits are also included in Fig. 2 and agree well with the data. A spin depolarization time ($\tau_1$) for trions (excitons) of ~103 ps (158 ps), is extracted for this specific gate voltage. These values are consistent with previous measurements on QWs' [3]. These shorter lifetimes are tunable with the back-gate voltage (see Supplementary). The focus here is the longer time constant ($T_1^{2DEG}$), associated with the spin decay time of the 2DEG, measured as 850 ps (1050 ps) for probe photon energies at the trion (exciton), at this specific voltage. The presence of a ~ ns time constant for the (rather short-lived) exciton originates from the relaxation of the exciton to the trion state and the related initialization of 2DEG spin polarization [23,24].

divided into three distinct regimes, similar to the situation for the PL in Fig. 1(b). Below $V_B$ = -1 V, the spectral response of the exciton and trion overlaps. Consequently, the extracted time-scales for both probe photon energies are similar. As $V_B$ is tuned to the range between -1 V and 0 V, the $T_1^{2DEG}$ times at trion and exciton probe energies increase drastically (~ 3 times) compared to more negative bias. This finding is related to an enhancement of the trion formation as the electron concentration of the 2DEG reaches an optimal value [14,25]. The increased spin decay time is also related to a reduced influence of the nuclear field fluctuations, which often affect the low-temperature electron spin relaxation in QWs [23,26]. The spin decay times for detection at the exciton and trion resonance differ in this intermediate bias regime. Specifically, $T_1^{2DEG}$ for probe tuned to the exciton energy, reaches its maximum at $V_B$ = – 0.35 V. For probe tuned to trion energy, $T_1^{2DEG}$ is maximum at a slightly different bias of $V_B$ = – 0.2 V. The small difference between voltages at which the longest $T_1^{2DEG}$ occurs for excitons and trions is related to the optimization of relative populations of excitons and trions. Interestingly, the longest $T_1^{2DEG}$ for probe energy at the trion is ~ 1.3 ns, whereas it is limited to ~ 1 ns for a probe tuned to the exciton. This is related to efficient initialization of the 2DEG by the trion.

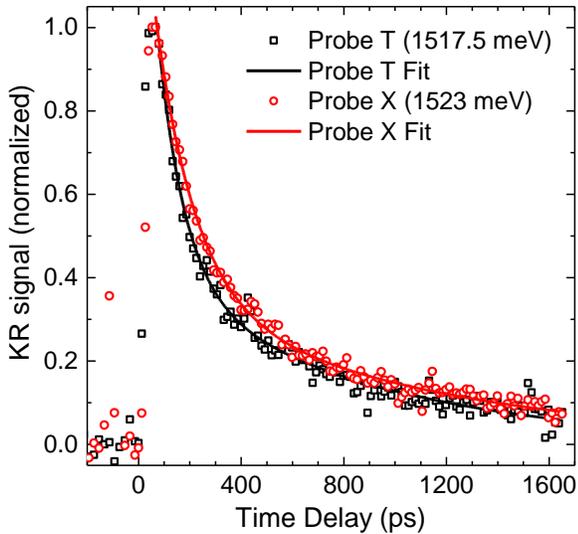

FIG. 2. KR transients detected at the trion (T) and exciton (X) for a back-gate voltage of $V_B$ = 0 V. The pump photon energy is 1.55 eV. Solid lines are bi-exponential fits.

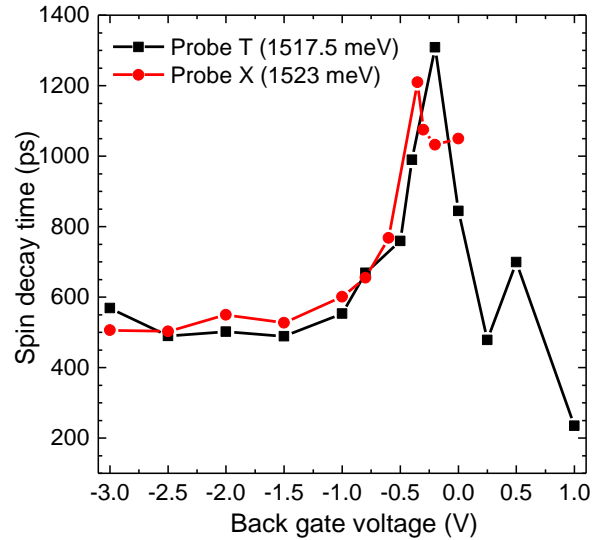

FIG. 3: Dependence of the spin decay time of the 2DEG ($T_1^{2DEG}$) on the back-gate voltage.

To further elaborate on the influence of the back-gate voltage (i.e., the electron density) on the long time scale ($T_1^{2DEG}$), we fit KR transients, using eq. (1), for a range of voltages and display the results in Fig. 3. The dependence of spin decay time on the voltage can be

The central result so far is that the spin decay time of the 2DEG is enhanced for an intermediate carrier density where excitons and trions coexist in the system. The non-monotonic dependence of spin lifetimes result from two competing mechanisms – nuclear field induced dephasing and Dyakanov-Perel mechanism, which have opposite dependence on carrier density [12]. When the gate voltage is further increased towards positive voltages, $T_1^{2DEG}$ is found to markedly decrease again for probing both excitons and trions. This voltage regime corresponds to a reduction in the trion population due to the screening effect of excess electrons in the QW, as also seen in the PL [27]. The reduction of the spin decay time for probe fixed to trion energy is even more drastic for voltages above 0 V. Corresponding data for probe tuned to the exciton is not shown as the 2DEG spin is not initialized through the exciton, as we demonstrate below.

Further evidence for inefficient spin initialization of the 2DEG through excitons, for positive voltages, is demonstrated via measurement of spin dynamics under the influence of an in-plane magnetic field (Voigt geometry), i.e., time-resolved MOKE. Specifically, we apply a magnetic field $B$ of ~ 227 mT, which causes spin precessions of the 2DEG. MOKE transients for two exemplary voltages of $V_B$ = -0.5 V and +0.5 V, for probe fixed at exciton energy, are displayed in Fig. 4. For $V_B$ = +0.5 V, the MOKE signal almost completely decays within ~ 250 ps, providing direct evidence for only minor spin initialization of the 2DEG at positive voltages. This finding originates from a reduction of the exciton-to-trion formation at these back-gate voltages [28]. The large number of electrons in the well screens the electron-hole interactions, making the formation of bound trions unfavorable [25,27]. As a result, the relaxation channel via trions is blocked, and 2DEG initialization is poor. In marked contrast, for $V_B$ = -0.5 V, the spin-related signals persist beyond our measurement range of ~1.7 ns, indicating coexistence of excitons and trions and resulting in an enhanced 2DEG spin initialization and lifetime.

The application of an in-plane magnetic field also allows us to extract the Larmor precession frequency and the transverse electronic g-factor from the MOKE transients. To this end, the oscillating part of the MOKE trace is fitted to a damped harmonic function of the form

$$\frac{dR}{R} = A_3 * e^{-\frac{t-t_0}{T_2^*}} * \sin(\omega t + \varphi) \quad (2),$$

where $T_2^*$ is the spin dephasing time (of the 2DEG) in an applied magnetic field, $\omega$ is the Larmor frequency of the 2DEG. Examples for such experimental data and corresponding fit results are shown in Fig. 5, for a probe tuned to the exciton resonance and $V_B$ = 0 V. Note that we have excluded the initial 300 ps from fitting, as this period is associated, e.g., with the spin relaxation of the trion. The spin dephasing time of the trion is likely shorter than a full Larmor period. Consequently the spin of the trion cannot complete a full precession around the external magnetic field. The later part of the MOKE signal reveals the Larmor precession of the 2DEG. It allows precise determination of the electron g-factor using $\hbar\omega = \mu_B g_B$, where $\mu_B$ is the Bohr magneton. The absolute values of the electron |g|-factors extracted from the MOKE signals (for probe tuned to the exciton energy), for the entire voltage range, are summarized in the inset of Fig. 5.

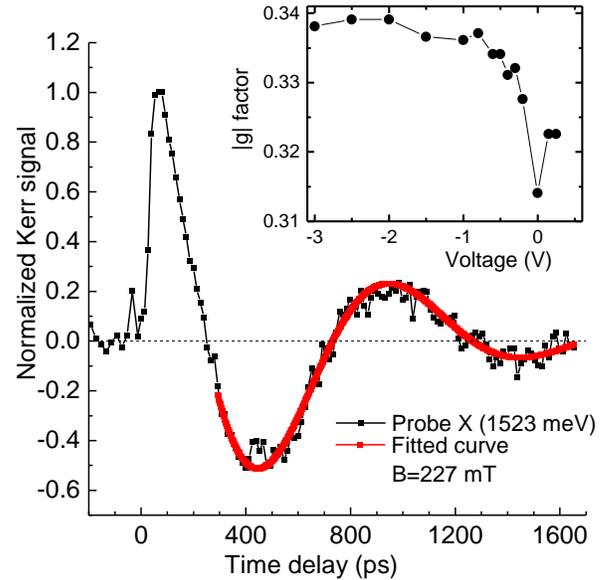

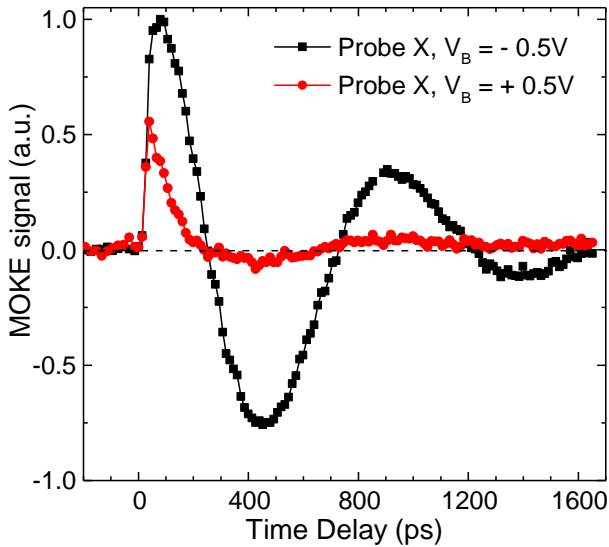

FIG. 4: MOKE transients for two exemplary back-gate voltages where trions co-exists or do not co-exist with the exciton. The probe is tuned to the exciton resonance (X).

FIG. 5 MOKE dynamics at $V_B$ = 0V, for probe fixed to exciton energy, and the curve fitting after 300 ps. Inset: voltage dependence of the absolute value of the extracted electron g-factor.

The absolute value of the electron |g|-factor (|g| ~ 0.33 at $V_B$ = -3 V, similar to values known from literature) are seen to decrease upon increasing the gate voltage towards positive values, i.e. with increasing density of carriers in the QW [23]. Given the g-factor of -0.44 for electrons in bulk GaAs and the observed trend towards positive values in narrow

QWs [29-32], we consider that the actual electron g-factor in our experiments is negative due to the wide width (20 nm) of the QW. Consequently, the trend of the inset of the Fig. 5 is flipped in sign, i.e., the electron g-factor changes towards more positive values when raising the back-gate voltage. This trend correlates well with theoretical calculations using *k•p* perturbation theory, which predicts an increase of the electron g-factor with the electron density [12]. The increasing of the electron g-factor is associated with a transition of band electrons from localized to delocalized states as carrier density increases. We note that the voltage dependence of the electron |g|-factor measured for probing at the trion position (data not presented here) follows the same trend as seen in the inset of Fig. 5.

## IV. CONCLUSIONS

In summary, we have shown that the PL, KR and the MOKE signals of the 20 nm GaAs QW are strongly dependent on the back gate voltage. An enhancement of the 2DEG spin lifetime for a particular range of the back gate voltages (from -1 V to 0 V), is observed. This coincides with the intermediate concentration of the charge carries in the QW predisposed to the coexistence of excitons and trions in the system, which in turn leads to an enhanced initialization of the 2DEG spin. The Larmor frequency associated to the MOKE signals is tuned via the electric field, correlating with an increase of the electron g-factors with the increasing carrier concentration and delocalization in the QW.

## ACKNOWLEDGMENTS

We acknowledge funding from the Deutsche Forschungsgemeinschaft (DFG) in the framework of the ICRC – TRR 160. Work at UT-Austin is supported by NSF DMR 1306878 and the Welch Foundation via grant F-1662. We thank Alan D. Bristow and Claudia Ruppert for help with the optical setup.

[1] T. C. Damen, L. Via, J. E. Cunningham, J. Shah, and L. J. Sham, *Subpicosecond spin relaxation dynamics of excitons and free carriers in GaAs quantum wells*, Phys. Rev. Lett. **67**, 3432 (1991).
[2] X. Marie, T. Amand, P. Le Jeune, M. Paillard, P. Renucci, L. E. Golub, V. D. Dymnikov, and E. L. Ivchenko, *Hole spin quantum beats in quantum-well structures*, Phys. Rev. B **60**, 5811 (1999).
[3] E. Vanelle, M. Paillard, X. Marie, T. Amand, P. Gilliot, D. Brinkmann, R. Levy, J. Cibert, and S. Tatarenko, *Spin coherence and formation dynamics of charged excitons in CdTe/CdMgZnTe quantum wells*, Phys. Rev. B **62**, 2696 (2000).
[4] A. Singh, G. Moody, K. Tran, M. E. Scott, V. Overbeck, G. Berghäuser, J. Schaibley, E. J. Seifert, D. Pleskot, N. M. Gabor, J. Yan, D. G. Mandrus, M. Richter, E. Malic, X. Xu, and X. Li, *Trion formation dynamics in monolayer transition metal dichalcogenides*, Phys. Rev. B **93**, 041401(R) (2016).
[5] P. Kossacki, V. Ciulin, M. Kutrowski, J.-D. Ganie, T. Wojtowicz, and B. Deveaud, *Formation Time of Negatively Charged Excitons in CdTe-Based Quantum Wells*, Phys. Stat. Sol. (b) **229**, 659 (2002).
[6] S. Bar-Ad and I. Bar-Joseph, *Exciton spin dynamics in GaAs heterostructures*, Phys. Rev. Lett. **68**, 349 (1992).
[7] A. S. Bracker, E. A. Stinaff, D. Gammon, M. E. Ware, J. G. Tischler, A. Shabaev, A. L. Efros, D. Park, D. Gershoni, V. L. Korenev, and I. A. Merkulov, *Optical pumping of the electronic and nuclear spin of single charge-tunable quantum dots*, Phys Rev Lett **94**, 047402 (2005).
[8] J. Tribollet, F. Bernardot, M. Menant, G. Karczewski, C. Testelin, and M. Chamarro, *Interplay of spin dynamics of trions and two-dimensional electron gas in an-doped CdTe single quantum well*, Physical Review B **68** (2003).
[9] M. V. Dutt, J. Cheng, B. Li, X. Xu, X. Li, P. R. Berman, D. G. Steel, A. S. Bracker, D. Gammon, S. E. Economou, R. B. Liu, and L. J. Sham, *Stimulated and spontaneous optical generation of electron spin coherence in charged GaAs quantum dots*, Phys Rev Lett **94**, 227403 (2005).
[10] G. Finkelstein, H. Shtrikman, and I. Bar-Joseph, *Optical Spectroscopy of a Two-Dimensional Electron Gas near the Metal-Insulator Transition*, Phys. Rev. Lett. **74**, 976 (1995).
[11] R. Bratschitsch, Z. Chen, S. T. Cundiff, E. A. Zhukov, D. R. Yakovlev, M. Bayer, G. Karczewski, T. Wojtowicz, and J. Kossut, *Electron spin coherence in n-doped CdTe/CdMgTe quantum wells*, App. Phys. Lett. **89** (2006).
[12] Z. Chen, S. G. Carter, R. Bratschitsch, and S. T. Cundiff, *Optical excitation and control of electron spins in semiconductor quantum wells*, Physica E: Low-dimensional Systems and Nanostructures **42**, 1803 (2010).
[13] T. C. Damen, K. Leo, J. Shah, and J. E. Cunningham, *Spin relaxation and thermalization of excitons in GaAs quantum wells*, App. Phys. Lett. **58**, 1902 (1991).
[14] G. Yusa, H. Shtrikman, and I. Bar-Joseph, *Onset of exciton absorption in modulation-doped GaAs quantum wells*, Phys. Rev. B **62**, 15390 (2000).
[15] R. Dingle, H. L. Störmer, A. C. Gossard, and W. Wiegmann, *Electron mobilities in modulation-doped*

*semiconductor heterojunction superlattices*, App. Phys. Lett. **33**, 665 (1978).

[16]  D. Brinkmann, J. Kudrna, P. Gilliot, B. Honerlage, A. Arnoult, J. Cibert, and S. Tatarenko, *Trion and exciton dephasing measurements in modulation-doped quantum wells: A probe for trion and carrier localization*, Phys. Rev. B **60**, 4474 (1999).

[17]  M. Z. Maialle, E. A. de Andrada e Silva, and L. J. Sham, *Exciton spin dynamics in quantum wells*, Phys. Rev. B **47**, 15776 (1993).

[18]  J. M. Kikkawa and D. D. Awschalom, *Resonant spin amplification in n-type GaAs*, Phys. Rev. Lett. **80**, 4313 (1998).

[19]  S. A. Crooker, D. D. Awschalom, J. J. Baumberg, F. Flack, and N. Samarth, *Optical spin resonance and transverse spin relaxation in magnetic semiconductor quantum wells*, Physical Review B **56**, 7574 (1997).

[20]  A. Singh, G. Moody, S. Wu, Y. Wu, N. J. Ghimire, J. Yan, D. G. Mandrus, X. Xu, and X. Li, *Coherent Electronic Coupling in Atomically Thin $MoSe_2$*, Physical review letters **112** (2014).

[21]  K. Kheng, R. T. Cox, Y. Merle A., F. Bassani, K. Saminadayar, and S. Tatarenko, *Observation of negatively charged excitons $X^-$ in semiconductor quantum wells*, Phys. Rev. Lett. **71**, 1752 (1993).

[22]  G. Finkelstein, V. Umansky, I. Bar-Joseph, V. Ciulin, S. Haacke, J.-D. Ganiere, and B. Deveaud, *Charged exciton dynamics in GaAs quantum wells*, Phys. Rev. B **58**, 12637 (1998).

[23]  I. Y. Gerlovin, Y. P. Efimov, Y. K. Dolgikh, S. A. Eliseev, V. V. Ovsyankin, V. V. Petrov, R. V. Cherbunin, I. V. Ignatiev, I. A. Yugova, L. V. Fokina, A. Greilich, D. R. Yakovlev, and M. Bayer, *Electron-spin dephasing in $GaAs/Al_{0.34}Ga_{0.66}As$ quantum wells with a gate-controlled electron density*, Physical Review B **75** (2007).

[24]  T. A. Kennedy, A. Shabaev, M. Scheibner, A. L. Efros, A. S. Bracker, and D. Gammon, *Optical initialization and dynamics of spin in a remotely doped quantum well*, Physical Review B **73**, 045307 (2006).

[25]  D. Sanvitto, R. A. Hogg, A. J. Shields, M. Y. Simmons, D. A. Ritchie, and M. Pepper, *Formation and Recombination Dynamics of Charged Excitons in a GaAs Quantum Well*, physica status solidi (b) **227**, 297 (2001).

[26]  R. I. Dzhioev, V. L. Korenev, B. P. Zakharchenya, D. Gammon, A. S. Bracker, J. G. Tischler, and D. S. Katzer, *Optical orientation and the Hanle effect of neutral and negatively charged excitons in $GaAs/Al_xGa_{1-x}As$ quantum wells*, Physical Review B **66** (2002).

[27]  A. J. Shields, M. Pepper, D. A. Ritchie, M. Y. Simmons, and G. A. C. Jones, *Quenching of excitonic optical transitions by excess electrons in GaAs quantum wells*, Physical Review B **51**, 18049 (1995).

[28]  M. T. Portella-Oberli, J. Berney, L. Kappei, F. Morier-Genoud, J. Szczytko, and B. Deveaud-Pledran, *Dynamics of Trion Formation in $In_xGa_{1-x}As$ QuantumWells*, Phys. Rev. Lett. **102**, 096402 (2009).

[29]  M. J. Snelling, E. Blackwood, C. J. McDonagh, R. T. Harley, and C. T. B. Foxon, *Exciton, heavy-hole, and electron g factors in type-I $GaAs/Al_xGa_{1-x}As$ quantum wells*, Physical Review B **45**, 3922 (1992).

[30]  M. J. Snelling, G. P. Flinn, A. S. Plaut, R. T. Harley, A. C. Tropper, R. Eccleston, and C. C. Phillips, *Magnetic g factor of electrons in $GaAs/Al_xGa_{1-x}As$ quantum wells*, Physical Review B **44**, 11345 (1991).

[31]  P. L. Jeune, D. Robart, X. Marie, T. Amand, M. Brousseau, J. Barrau, V. Kalevich, and D. Rodichev, *Anisotropy of the electron Landé g factor in quantum wells*, Semiconductor Science and Technology **12**, 380 (1997).

[32]  R. M. Hannak, M. Oestreich, A. P. Heberle, W. W. Rühle, and K. Köhler, *Electron g factor in quantum wells determined by spin quantum beats*, Solid State Communications **93**, 313 (1995).